\documentstyle[icmp,epsfig]{article}

\input epsf

\def\Journal#1#2#3#4{{#1} {\bf #2}, #3 (#4)}

\def\be{\begin{equation}}
\def\ee{\end{equation}}
\def\bea{\begin{eqnarray}}
\def\eea{\end{eqnarray}}

\begin{document}
\title{SYSTEMATICS OF THE GENETIC CODE AND ANTICODE:
HISTORY, SUPERSYMMETRY, DEGENERACY AND PERIODICITY}

\author{P D JARVIS, J D BASHFORD\footnote{
Department of Physics and Mathematical Physics, 
University of Adelaide \\ E-mail: jbashfor@physics.adelaide.edu.au}}

\address{School of Mathematics and Physics, University of Tasmania \\ 
E-mail: peter.jarvis@utas.edu.au } 

\maketitle\abstracts{}

\vspace*{-.5cm}
The evolution of life is thought to have 
proceeded via an `RNA world' with RNA as the 
functional and information storage
medium. The original role for peptides was to stabilise
RNA enzymes (ribozymes) \cite{Eigen}.
The translation of nucleic acids into proteins occurs via
$t$RNA adaptor molecules which carry the amino acids for
assembly. The $t$RNA's get charged with their specific
amino acid in interaction with an amino acyl-$t$RNA
synthetase,
which recognises the $t$RNA anticodon and the appropriate amino acid
(phylogenetic studies \cite{Ohno} have suggested complementary
relationships between the
two classes of $t$RNA synthetases, which
use two different amino acid attachment sites, and 
similarly between the $t$RNA's themselves).
The dynamical basis of algebraic approaches to the genetic code lies,
in biochemical terms, in the complementary 
bonding properties of nucleoside bases in $m$RNA codons and in $t$RNA
anticodons, and also in the affinities of $t$RNA species for particular
amino acids via conformational interaction between the 
sites of anticodon binding and amino-acylation during charging. 
If it is assumed that code evolution
represents an optimisation process, there is hope for an account 
in terms of broken dynamical symmetries \cite{Hornos,Others}. 
In such schemes, it can be expected that
correlations
between physico-chemical properties of anticodons and amino acids
should reflect their organisation under various subalgebra 
chains; furthermore,   
the details of branching schemes at the level of individual
weight vectors should be consistent with accepted biological understandings
of code history. 

\begin{figure}[htb]

\hspace*{0.8cm}
\epsfig{angle=90,figure=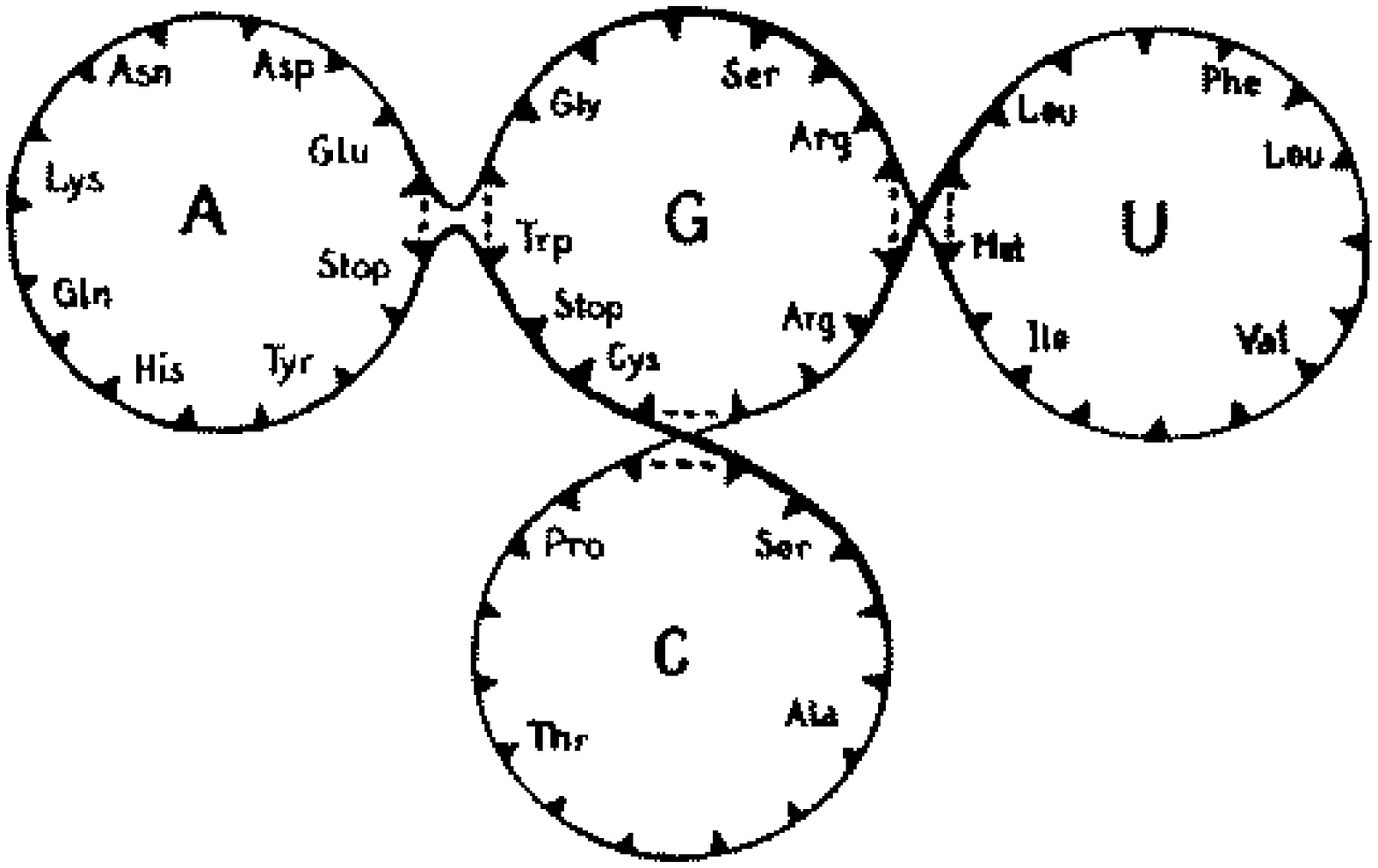,height=6cm}
\caption{One step mutation rings for the genetic code (after Siemion)}
\label{fig:rings}
\end{figure}

A recent model \cite{Us} based on 64-dimensional typical 
irreducible representations of the classical Lie superalgebra
$sl(6/1)$ fulfils these requirements in that the branching diagram mimics
the well-known tabular presentation of the code. For example, 
the dominance of the middle base letter in determining anticodon
hydrophilicity (with the $-A-$ and $-U-$ families showing most 
divergence, and the $-C/G-$ families intermediate) corresponds to
a first branching stage wherein $64_b \rightarrow 1\times16_b
+ 2\times 16_{b+1} + 1 \times 16_{b+2}$ with respect to $sl(2)+sl(4/1)$,
with $-A-$ and $-U-$ as singlets, and magnetic splitting between the $-C-$
and 
$-G-$ doublet only 
implemented at a later stage\footnote{The subscript refers to the single nonzero
(real number) Dynkin label of the $2^n$ dimensional typical irrep of 
$sl(n/1)$}. The above points are also addressed in our model
via Siemion's studies \cite{Siemion} of the periodicity of the 
genetic code. The analogue of atomic number for the genetic code is
organisation of codons into abstract `one step mutation
rings' where the sequence
of shifts {\sf 333-1-333-1-333-1-333-2-$\cdots$} is listed,
with the `mutations' on the indicated base letter position being 
$U \rightarrow C \rightarrow G \rightarrow A$ repeated cyclically
or anticyclically after each higher level substitution.
The three excursions into different second bases can be
arranged so that the structure has the topology of four
interlocking
rings, with $C$ and $G$ central, and $A$ and $U$ outlying,
as in the discussion of hydrophilicity (see figure \ref{fig:rings}).
There is a solid body of evidence that this arrangement 
exhibits periodicity in a host of experimental measures. We end by 
demonstrating \cite{Us}
how the figure can be regarded as a projection of the weight 
diagram of the 64-dimensional codon irrep of $sl(6/1)$.

\begin{figure}[htb]
\epsfig{angle=90,figure=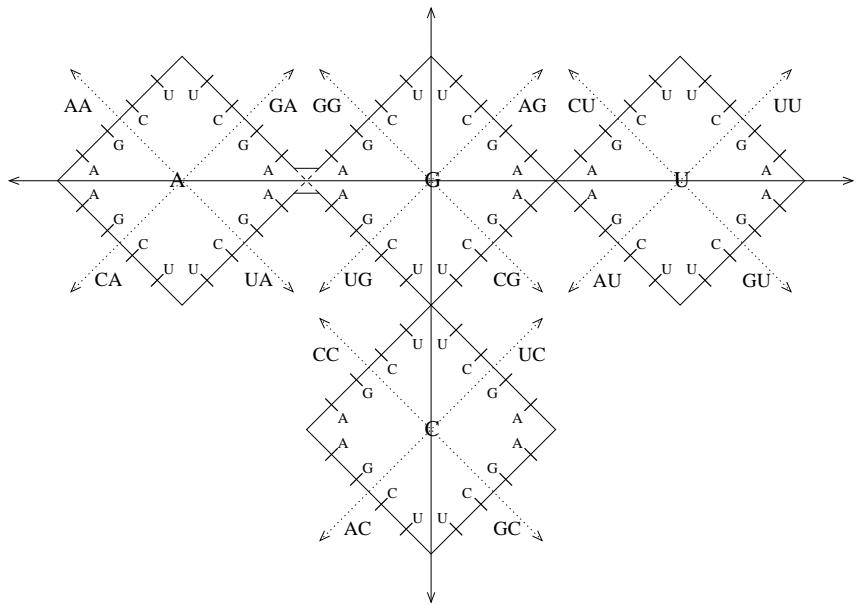,height=8cm}
\caption{Concatenated weight diagram of the 64 dimensional irrep of 
$sl(6/1)$}
\label{fig:diamonds}
\end{figure}

Firstly, it is clear that the four ring pattern is convincingly similar to
the above branching rule for the 64 into hexadecuplets of $sl(4/1)$, 
provided the 
plot \footnote{The parameters $\Delta$ represent the respective
the Dynkin label shifts, and take values $-1,~0,~+1$; superscripts refer to
base position within codons} is of
`hypercharge' $Y \equiv {J_3}^{(2)}-\frac 12$ versus $\Delta^{(2)}$.
This same pattern repeats for each 16, with each ring
being thought of as a small weight diagram of ${J_3}^{(1)} ~vs~ \Delta^{(1)}$ 
superimposed on the $Y ~ vs ~ \Delta^{(2)}$ plane, on which  
four family boxes (quartets of $sl(2/1)$) are displayed. Finally, 
individual codons (and amino acids) are located by a
{\it one-dimensional} projection
of these quartets with respect to $\Delta^{(3)} + {J_3}^{(3)}$. The 
orientations of each of these plots is determined by position on the rings, 
and the weights are 
arrayed as a diamond rather than a circle. The result is shown in 
figure \ref{fig:diamonds}.

Siemion's periodicity of the ring 
structure (as one example \cite{Siemion}, with respect to the 
Chou-Fasman amino acid $P_\beta$ conformational parameters,
which give statistical information on amino acid usages in
protein $\beta$ sheets) is expected to be reproduced by dynamical symmetry
operators which depend on the above weight labels and associated Casimirs 
in a consistent way; work along these lines is in progress. 
The vexed issue of 
code {\it degeneracy} in algebraic schemes is, in our model \cite{Us},
also 
related to the periodicity: identical amino acids are `captured' by 
$t$RNA's working on equivalent positions with respect to reflection 
symmetries of the ring pattern \cite{Siemion} (the functional dependence
of Casimir operators must be chosen to respect such symmetries). 
In the case of $Ser$, it is obvious for example that the 1$\frac 12$ 
family boxes responsible for its six-fold degeneracy occur on the $C$ 
and $G$ rings at roughly the {\it same} 
relative location.

\subsection*{Acknowledgements}
The authors would like to thank
Chris Jarvis for helpful correspondence.

\vfill
\pagebreak

\nopagebreak

\end{document}